\pdfoutput=1 

\documentclass{article} 
\usepackage{icrc2011}
\usepackage{units}

\title{Observations of Large Scale Sidereal Anisotropy in 1 and 11 TeV cosmic rays from the MINOS experiment.}

\newcommand{\etal}{\MakeLowercase{\textit{et al. }}} 
\shorttitle{de Jong \etal  Sidereal Anisotropies observed in MINOS}

\authors{J.~K.~de Jong$^{1}$ for the MINOS Collaboration }
\afiliations{$^1$ Department of Physics, Oxford University, Oxford, United Kingdom OX1 3RH }
\email{jeffrey.dejong@physics.ox.ac.uk}

\abstract{The MINOS Near and Far Detectors are two large, functionally-identical, steel-scintillating sampling calorimeters located at depths of 220~mwe and 2100~mwe respectively. The detectors observe the muon component of hadronic showers produced from cosmic ray interactions with nuclei in the earth's atmosphere. From the arrival direction of these muons, the anisotropy in arrival direction of the cosmic ray primaries can be determined. The MINOS Near and Far Detector have observed anisotropy on the order of 0.1\% at 1 and 11 TeV respectively. The amplitude and phase of the first harmonic at 1 TeV are 8.2$\pm$1.7(stat.)$\times 10^{-4}$ and (8.9$\pm$12.1(stat.))$^{\circ}$, and at 11 TeV are 3.8$\pm$0.5(stat.)$\times 10^{-4}$ and (27.2$\pm$7.2(stat.))$^{\circ}$.
}
\keywords{MINOS,~Atmospheric muons,~Seasonal Variations,~Sidereal Anisotropy }

\begin{document}
\maketitle
\section{Introduction}

Cosmic ray primaries in the 1~TeV to 100~TeV energy regime are thought to originate within our galaxy~\cite{berez}. Curvature in the galactic magnetic field~(GMF) and multiple interactions with local scattering objects are expected to eliminate point-of-origin information and cause the observed arrival direction of cosmic ray primaries to be highly isotropic.  However, large scale anisotropies at these energies have been observed by Milagro~\cite{milagro}, the Tibet Air Shower Array~\cite{tibet}, ARGO-YBJ~\cite{argo}, IceCube~\cite{icecube} and Super-Kamiokande~\cite{superk}.\\

 Anisotropy can be introduced through multiple different mechanisms. Compton and Getting~\cite{compton} predicted a dipole effect could arise due to the rotation of the solar system around the galactic center with the excess appearing in the direction of motion. Diffusion of cosmic rays out of the galactic-disk into the galactic-halo could introduce an anisotropy perpendicular to the disk~\cite{berez}. At TeV energies the heliosphere could produce a cosmic-ray excess in the direction of the heliotail; this excess should diminish with energy and is expected to be negligible above \unit[10]{TeV}~\cite{nagashima}. Discrete sources have been shown capable of creating anisotropy at higher energies~\cite{ptuskin,strong}. \\

MINOS is a two detector long baseline neutrino beam oscillation experiment\cite{minos}. The \unit[0.98]{kton} Near Detector, located at the Fermi National Accelerator Laboratory in Batavia, IL., is situated \unit[94]{m} underground, below a flat overburden of 225 meters of water equivalent~(mwe). The vertical muon threshold is \unit[0.05]{TeV}. The raw atmospheric muon trigger rate is \unit[27.29$\pm$0.01]{Hz}.  The \unit[5.4]{kton} Far Detector is located in the Soudan Underground Mine State Park in Soudan, MN., \unit[720]{m} below the surface. The vertical overburden is \unit[2100]{mwe} and the muon threshold is \unit[0.73]{TeV}. The trigger rate at the Far Detector is \unit[0.5364$\pm$0.0001]{Hz}. The MINOS detectors are two large, functionally-identical, steel-scintillating sampling calorimeters. The steel-scintillator planes are oriented vertically for optimal detection of horizontal beam neutrinos. As a consequence the detector has negligible acceptance for vertical muons increasing to near unity for horizontal muons.\\

The MINOS detectors do not measure the arrival direction or energy of the cosmic ray primary. These quantities are inferred from the muons produced in the air shower. These muons are highly boosted objects and travel nearly parallel to the direction of the original primary. The mean surface energy for muons observed at the Near(Far) Detector is \unit[0.1]{TeV}(\unit[1.1]{TeV}). The mean cosmic ray primary energy is ten times larger~\cite{murakami} than the mean muon surface energy. The Near and Far Detectors therefore measure the anisotropy of \unit[1.0]{TeV} and \unit[11]{TeV} primaries respectively.\\

The expected level of cosmic ray anisotropy at TeV energies is \unit[0.1]{\%}. Spurious sidereal anistropies can be created by the detectors' non-uniform sky exposure. The diurnal and seasonal variation in the muon flux due to variation in the atmospheric temperature can interfere to produce a anisotropy with a sidereal component. Both of these effects must be removed to obtain a reliable measurement of the true cosmic ray anisotropy. The former has be mitigated by accurately measuring the detector livetime; the latter by first measuring and then subtracting out the seasonal variation component.\\

\section{Event Selection}
\label{sec:Selection}

The goal of this analysis is to identify large scale anisotropies in the arrival direction of atmospheric muons in the MINOS data set. The size of the data-set limits the search to anisotropies larger than $10^{\circ}$. These muons travel near parallel to the direction of the original cosmic ray primary; the large underground momentum of the muons limits the amount of multiple scattering in the overburden to be no more than \unit[2]{$^{\circ}$}. The selection criteria have been chosen to obtain a detector pointing accuracy that, when combined with the scattering in the overburden and the initial scatter of the muon, gives an angular resolution on the direction of the primary cosmic-ray of 5$^{\circ}$. Moon shadow searches performed using the MINOS Far Detector data have demonstrated that the detectors have a negligible pointing bias~\cite{moon}.\\

The same selections are applied to the data collected at the Near and Far Detectors. The candidate event must contain one and only one track. This track must be longer than \unit[2]{m} and possess a vertex which is no further than \unit[50]{cm} from the detector edge. The $\chi^{2}/ndf$ of the track-fit must be less than 1, and the residual of the track to a straight line must be less than \unit[0.04]{m}. Finally, it is required that the data be collected during a period of good detector run conditions. In summary the Far~(Near) Detector had an effective live-time of \unit[93.4]{\%}~(\unit[76.0]{\%}), and a selection efficiency of \unit[63.8]{\%}~(\unit[32.4]{\%}). The Near Detector selection efficiency is much lower than that in the Far Detector as events in the down-stream end of the detector trigger the readout but are not reconstructed. After all selections we retain 67.7$\times 10^{6}$ events in the Far Detector and 0.989$\times 10^{9}$ events in the Near Detector.\\

\section{Seasonal Variations}
\label{sec:Seasonal}
When cosmic ray primaries interact with nuclei in the upper atmosphere the subsequent showers contains kaons and pions. These mesons can either interact or decay producing the muons which we observe underground. The probability that a meson will decay is proportional to its energy and the density of the atmosphere through which it travels. Increasing the temperature of the atmosphere decreases the density reducing the probability that a meson will interact thereby increasing the observed muon rate underground. The relationship between atmospheric temperature and muon rate underground can be written as
\begin{equation}
\frac{\Delta R_{\mu}}{\left<R_{\mu}\right>}=\alpha_{T}\frac{\Delta T_{\mathrm{eff}}}{\left< T_{\mathrm{eff}}\right>}
\label{eq:eqn1}
\end{equation}
where $\Delta R_{\mu}$ is the deviation from the overall average muon rate. $\left<R_{\mu}\right>$ is the muon intensity evaluated at a temperature $\left<T_{\mathrm{eff}}\right>$. $\alpha_{T}$ is the coefficient of correlation between rate and temperature. The temperature T$_{\mathrm{eff}}$ is an effective temperature of  the atmosphere which weights the temperature of the atmosphere at different depths by the probability of a muon being produced at that elevation~\cite{season1}. The temperature as a function of depth has been determined using the European center for Medium-Range Weather Forecasts(ECMWF) atmospheric model~\cite{ecmwf}. The model provides atmospheric temperatures at 37 different pressure levels between 1 and \unit[1000]{hPa} at four different times (\unit[0000]{h}, \unit[0600]{h}, \unit[1200]{h},\unit[1800]{h}) throughout the day. \\

\begin{figure}
\begin{center}
\includegraphics[width=0.45\textwidth]{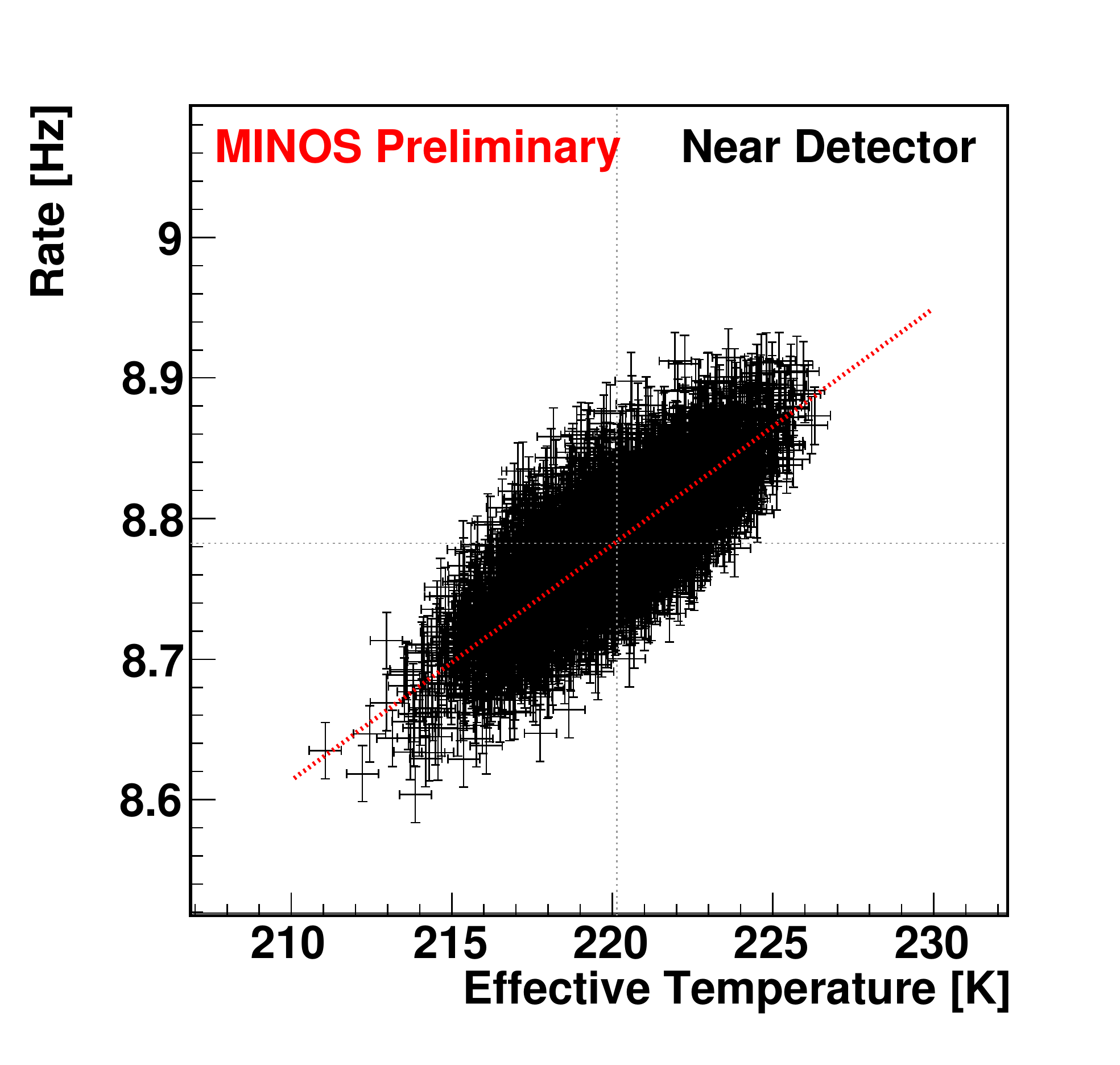}
\end{center}
\caption{The observed muon rate at the MINOS Near Detector versus effective atmospheric temperature. The data is binned in \unit[6]{hr} intervals centered around the ECMWF data points. The errors on the muon rate are statistical. The temperature errors are set to $\pm$0.5K. The red line is the best-fit of the data to equation~\ref{eq:eqn1}.}
\label{fig:Seasonal}
\end{figure}

The average muon rate and temperature at the MINOS Near~(Far) Detectors is 8.7820$\pm$0.003~Hz (0.32566$\pm$0.00004~Hz) and 220.1$\pm$0.5K (222.2$\pm$0.5K) respectively. Figure \ref{fig:Seasonal} plots the observed muon rate at the Near Detectors versus the effective atmospheric temperatures.  To determine the value of $\alpha_{T}$ a linear regression was performed on the data using ROOT's MINUIT fitting package. The best fit values for $\alpha_{T}$ are 0.421$\pm$0.004(stat.) for the Near Detector, and 0.886$\pm$0.009(stat.) for the Far Detector~\footnote{This measure of $\alpha_{T}$ at the Far Detector is in agreement with our previous publication~\cite{season1}}. The value of $\alpha_{T}$ is larger at the deeper Far Detector as it samples from higher energy mesons which have an increased probability of interacting.\\

\section{Sidereal Anisotropy}
\label{sec:Sidereal}

Detector acceptance effects dominate the number of events observed in declination making a true two-dimensional map, accurate to 0.1\%, impossible. The rotation of the earth washes out these effects in right ascension. If no anisotropy exists, and provided we account for differing exposure as a function of detector right ascension\footnote{Detector right ascension is the location to which the vertical axis of the detector is pointing}, we would expect the number of observed events to be flat as a function of right ascension.\\

The rate-weighted number of muons as a function of right-ascension ($\alpha$) and declination ($\delta$) are determined in each 1$^{\circ}$ detector right ascension bin $k$
\begin{equation}
n(\alpha,\delta)_{k}=\sum_{m=1}^{N_{\mu}}1(\alpha^{m},\delta^{m})\frac{<R>}{R(T_{\mathrm{eff}}^{m})}
\end{equation}
where $<R>$ is the average muon rate and R is the expected muon rate at the current effective temperature T$_{eff}^{m}$. The two-dimensional sky-map is determined by summing over all detector right ascension bins and correcting for the different live times $t_{K}$ in each bin.
\begin{equation}
N(\alpha,\delta)=\sum_{k=1}^{360}n(\alpha,\delta)_{k}\frac{<t>}{t_{K}}
\end{equation}
where $\left< t \right>$ is the average detector live time per detector right ascension bin. The level of anisotropy $A$ in the right-ascension bin $\alpha_{i}$ for a given declination bin $\delta_{j}$ is
\begin{equation}
A(\alpha_{i},\delta_{j})=\frac{N(\alpha_{i},\delta_{j})-<N(\delta_{j})>}{<N(\delta_{j})>}
\end{equation}
where $<N(\delta_{j})>$ is the average number of events per right ascension bin for that declination bin. The one-dimensional sky-map is obtained by integrating the data over all declination bins.\\

\begin{figure*}[h!t!b]
\begin{center}
\includegraphics[width=0.4\textwidth]{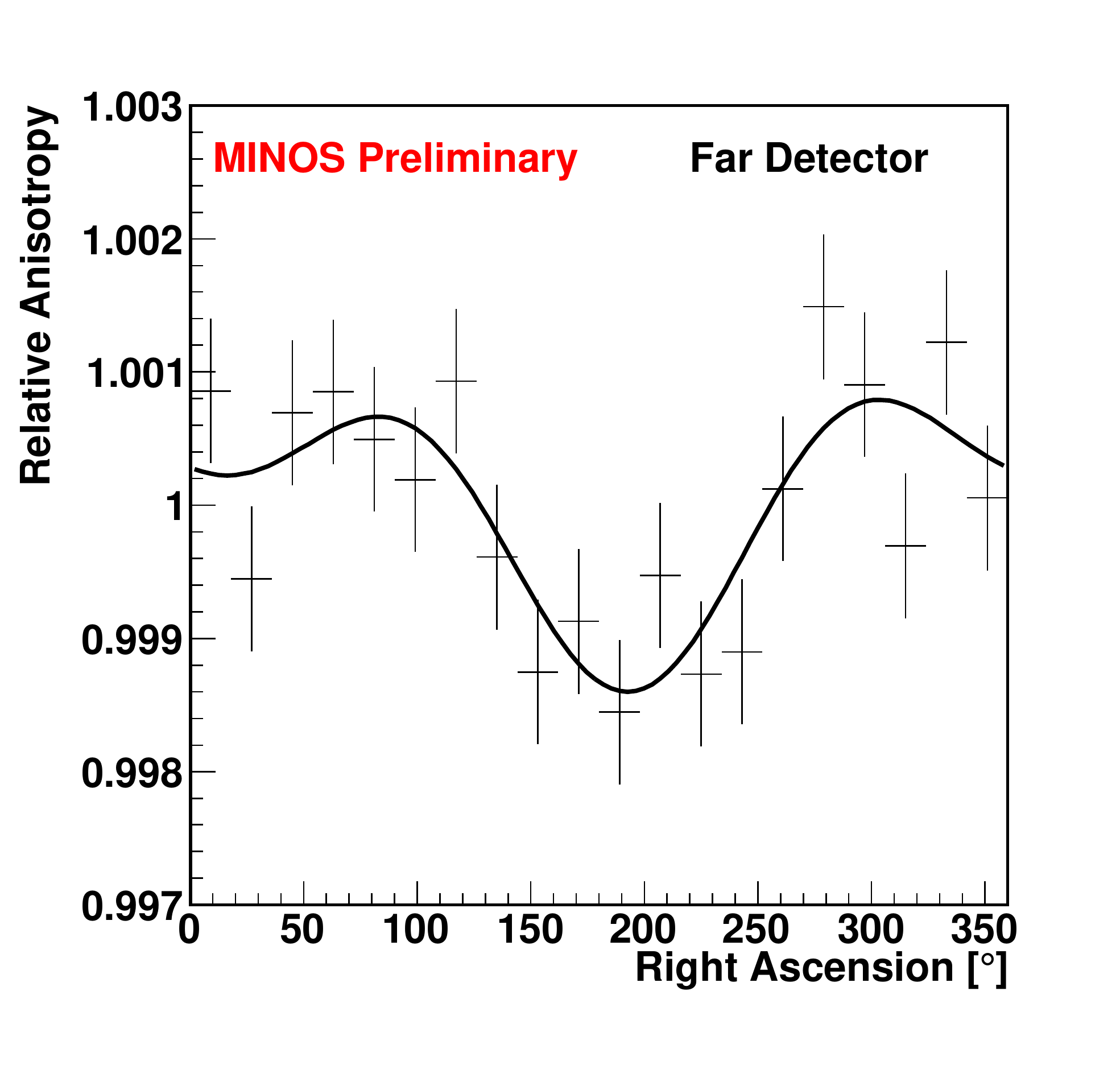}
\includegraphics[width=0.4\textwidth]{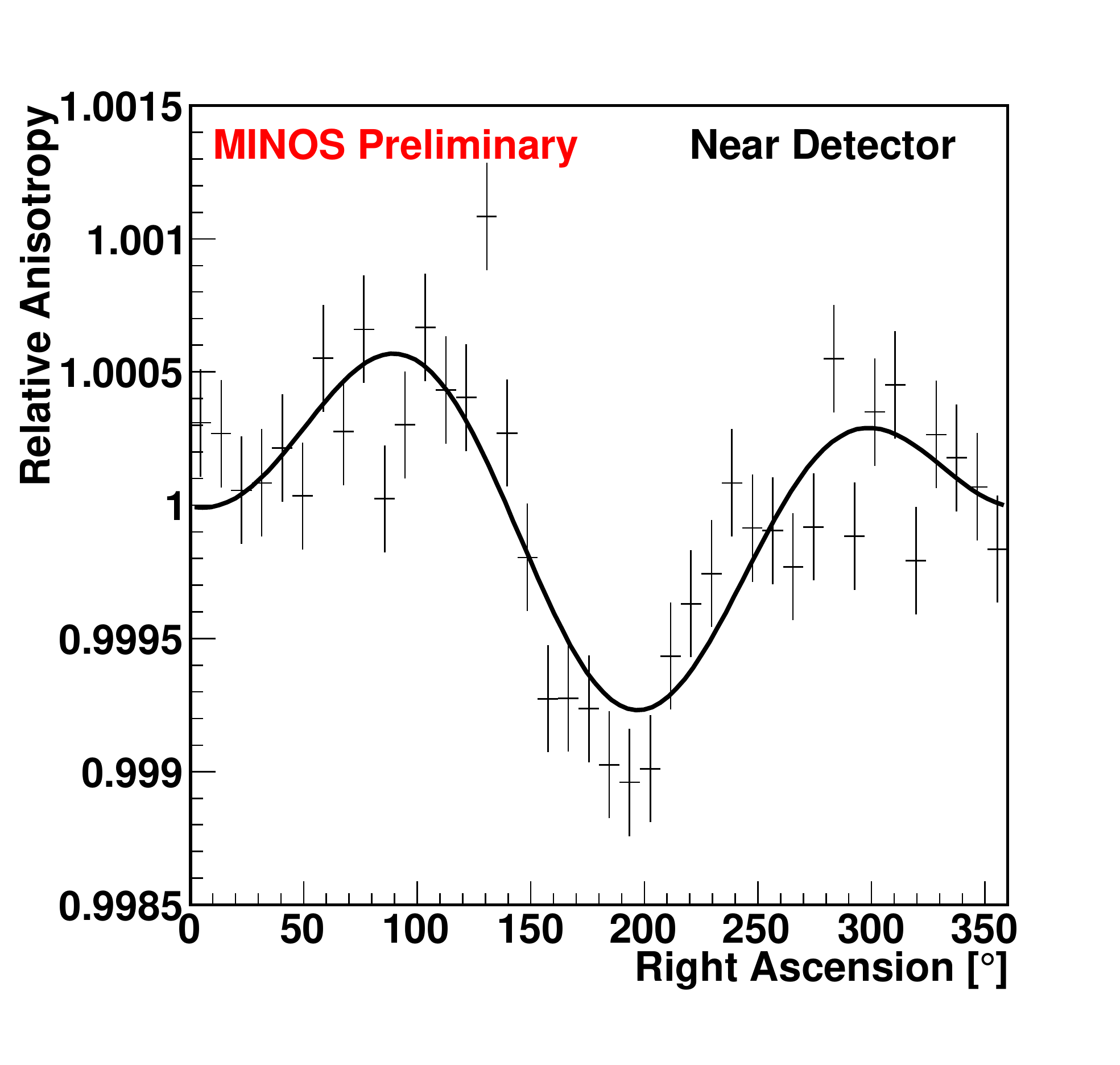}
\end{center}
\caption{The one-dimensional projection of the Near and Far Detector sky-maps. The level of anisotropy in both data samples is 0.1\%. The mean primary energy for the Far Detector is \unit[11]{TeV}, and for the Near Detector is \unit[1]{TeV}.}
\label{fig:FD1D}
\end{figure*}

A second order harmonic is typically used to describe the one-dimensional anisotropy
\begin{equation}
A(\alpha)=1+\sum_{n=1}^{2}A_{n}cos\left[\frac{n\pi}{180}(\alpha-\phi_{n}) \right].
\label{eq:bestfit}
\end{equation}
Figure \ref{fig:FD1D} plots the one-dimensional cosmic-ray sidereal anisotropy as  observed by the atmospheric muons observed in the MINOS detectors. The data in figure \ref{fig:FD1D} was fit to equation~\ref{eq:bestfit}. The amplitude and phase of the primary harmonic are 3.8$\pm$0.5(stat.)$\times 10^{-4}$ and 27.2$\pm$7.2(stat)$^{\circ}$ for the Far Detector data~(\unit[11]{TeV} primary) and 8.2$\pm$1.7(stat.)$\times 10^{-4}$ and a phase of 8.9$\pm$12.1(stat)$^{\circ}$ for the Near Detector data~(\unit[1]{TeV} primary).  A \unit[5.3]{$\sigma$} deficit is observed in the Far detector data between 150$^{\circ}$ and 245$^{\circ}$, the chance probability of this deficit occurring randomly is 0.001\%. An \unit[11.1]{$\sigma$} deficit is observed in the Near Detector data between 155$^{\circ}$ and 225$^{\circ}$ with a chance probability much less than 0.0001\%. The Near detector also sees a \unit[7.8]{$\sigma$} excess between 50$^{\circ}$ and 140$^{\circ}$ also with a chance probability much less than 0.0001\%.\\

\begin{figure*}[h!b!t!]
\begin{center}
\includegraphics[width=0.8\textwidth]{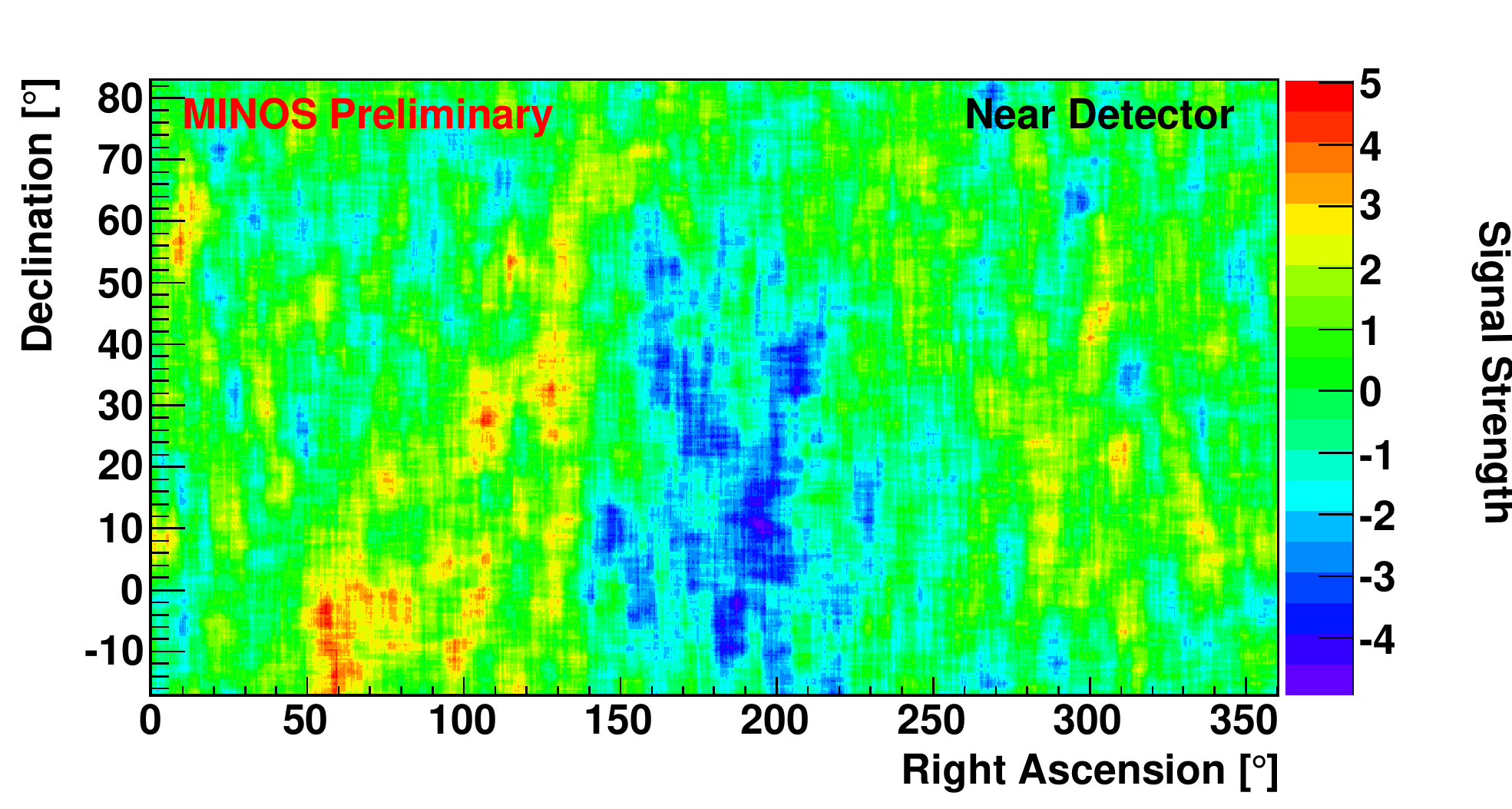}
\end{center}
\caption{The two-dimensional significance sky-map for \unit[1]{TeV} cosmic rays as determined using the MINOS Near Detector. A $\unit[5]{^{\circ}}$ smoothing has been applied.}
\label{fig:ND2D}
\end{figure*}

The large statistics collected using the Near Detector are sufficient to allow an exploration of the two-dimensional anisotropy sky-map. Figure \ref{fig:ND2D} plots the two-dimensional significance, defined as
\begin{equation}
\sigma(\alpha_{i},\delta_{j})=\frac{N(\alpha_{i},\delta_{j})-<N(\delta_{j})>}{\sqrt{<N(\delta_{j})>}}
\end{equation}
for \unit[1]{TeV} CR primaries, where $<N(\delta_{j})>$ is the total number of events in the 5$^{\circ}$ smoothed data sample. A significant deficit is observed Near the north galactic pole (192$^{\circ}$ ra, +27$^{\circ}$ dec), and an excess is observed around the helio-tail. The energy of our Near Detector data sample is equivalent to that collected by ARGO-YBJ experiment~\cite{argo}; and the two-dimensional sky-maps look similar.

\section{Conclusion}
The MINOS Near and Far Detector have been used to measure the anisotropy in the arrival direction of cosmic rays at energies of \unit[1]{TeV} and \unit[11]{TeV}. The amplitude of the first-harmonic has been observed to decrease with increasing energy from 8.2$\pm$1.7(stat.)$\times 10^{-4}$ at \unit[1]{TeV} to 3.8$\pm$0.5(stat.)$\times 10^{-4}$ at  \unit[11]{TeV}. The two-dimensional sky-map at \unit[1]{TeV} is consistent with previous publications. The MINOS experiment continues to take data and a more detailed analysis of the data will be forthcoming.

\section{Acknowledgments}
This work was supported by the US DOE, the UK STFC, the US NSF, the State and University of Minnesota, the University of Athens, Greece and Brazil's FAPESP and CNPq. We are grateful to the Minnesota Department of Natural Resources, the crew of Soudan Underground Laboratory, and the staff of Fermilab for their contributions to this effort.

\end{document}